\documentclass[
preprint,
superscriptaddress,
amsmath,amssymb,
aps,
pre,
floatfix,
]{revtex4-2}

\usepackage{graphicx}
\usepackage{dcolumn}
\usepackage{bm}
\usepackage[
left=0.7in,right=0.7in,top=0.8in,bottom=0.9in
]{geometry}
\usepackage{subfigure}
\usepackage{latexsym}
\usepackage{epsfig}
\usepackage{amsbsy}
\usepackage{empheq}
\usepackage{cases}

\usepackage[utf8]{inputenc}
\usepackage[T1]{fontenc}

\usepackage{array}
\usepackage{setspace}
\usepackage{lineno}
\usepackage{xcolor}
\usepackage{adjustbox}
\usepackage{multirow}
\usepackage[colorlinks,linkcolor=blue,anchorcolor=blue,urlcolor  = blue,citecolor=blue]{hyperref}
\usepackage{ulem}
\usepackage{txfonts}
\usepackage[section]{placeins}

\usepackage{verbatim} 

\newlength{\figurewidth}
\begin{document}

\title[]{Transmission spectroscopy of CF$_4$ molecules in intense x-ray fields}
	\author{Rui Jin}\email{rui.jin@mpi-hd.mpg.de}
	\affiliation {Max-Planck-Institut für Kernphysik, Saupfercheckweg 1, 69117 Heidelberg, Germany}

	\author{Adam Fouda} 
    \affiliation {Chemical Sciences and Engineering Division, Argonne National Laboratory, Lemont, Illinois 60439 USA}
    \affiliation {Department of Physics, University of Chicago, Chicago, Illinois 60637, USA}
    
	\author{Alexander Magunia}
    \affiliation {Max-Planck-Institut für Kernphysik, Saupfercheckweg 1, 69117 Heidelberg, Germany}  
    
	\author{Yeonsig Nam} 
    \affiliation {Chemical Sciences and Engineering Division, Argonne National Laboratory, Lemont, Illinois 60439 USA}

	\author{Marc Rebholz}
    \affiliation {Max-Planck-Institut für Kernphysik, Saupfercheckweg 1, 69117 Heidelberg, Germany} 

	\author{Alberto De Fanis}
    \affiliation {European XFEL, Holzkoppel 4, 22869 Schenefeld, Germany} 

	\author{Kai Li}
    \affiliation {Chemical Sciences and Engineering Division, Argonne National Laboratory, Lemont, Illinois 60439 USA}

	\author{Gilles Doumy} 
	\affiliation {Chemical Sciences and Engineering Division, Argonne National Laboratory, Lemont, Illinois 60439 USA} 

	\author{Thomas M. Baumann}
    \affiliation {European XFEL, Holzkoppel 4, 22869 Schenefeld, Germany}
    
    \author{Michael Straub}
    \affiliation {Max-Planck-Institut für Kernphysik, Saupfercheckweg 1, 69117 Heidelberg, Germany} 
    
    \author{Sergey Usenko}
    \affiliation {European XFEL, Holzkoppel 4, 22869 Schenefeld, Germany}
    
	\author{Yevheniy Ovcharenko}
	\affiliation {European XFEL, Holzkoppel 4, 22869 Schenefeld, Germany}
 
	\author{Tommaso Mazza}
	\affiliation {European XFEL, Holzkoppel 4, 22869 Schenefeld, Germany}
 
    \author{Jacobo Monta{\~{n}}o}
    \affiliation {European XFEL, Holzkoppel 4, 22869 Schenefeld, Germany} 	
    
	\author{Marcus Ag{\aa}ker} 
	\affiliation {Department of Physics and Astronomy, Uppsala University, P.O. Box 516, SE-751 20 Uppsala, Sweden}
    \affiliation{MAX IV Laboratory, Lund University, P.O. Box 118, SE-22100 Lund, Sweden}
     
    \author{Maria Novella Piancastelli}
	\affiliation{Sorbonne Universit{\'e}, CNRS, UMR 7614, Laboratoire de Chimie Physique-Mati{\`e}re et Rayonnement, F-75005 Paris, France}
	
	\author{Marc Simon} 
    \affiliation{Sorbonne Universit{\'e}, CNRS, UMR 7614, Laboratoire de Chimie Physique-Mati{\`e}re et Rayonnement, F-75005 Paris, France}
    
	\author{Jan-Erik Rubensson} 
    \affiliation{Department of Physics and Astronomy, Uppsala University, P.O. Box 516, SE-751 20 Uppsala, Sweden}
    
	\author{Michael Meyer} 
	\affiliation {European XFEL, Holzkoppel 4, 22869 Schenefeld, Germany}
 
	\author{Linda Young}\email{young@anl.gov}
    \affiliation {Chemical Sciences and Engineering Division, Argonne National Laboratory, Lemont, Illinois 60439 USA}
    \affiliation {Department of Physics, University of Chicago, Chicago, Illinois 60637, USA}
    
	\author{Christian Ott}\email{christian.ott@mpi-hd.mpg.de}
    \affiliation {Max-Planck-Institut für Kernphysik, Saupfercheckweg 1, 69117 Heidelberg, Germany}
    
    \author{Thomas Pfeifer}\email{thomas.pfeifer@mpi-hd.mpg.de}
    \affiliation {Max-Planck-Institut für Kernphysik, Saupfercheckweg 1, 69117 Heidelberg, Germany}
	\normalem
	\date{\today}

\begin{abstract}
The nonlinear interaction of x-rays with matter is at the heart of understanding and controlling ultrafast molecular dynamics from an atom-specific viewpoint, providing new scientific and analytical opportunities to explore the structure and dynamics of small quantum systems.  At increasingly high x-ray intensity, 
the sensitivity of ultrashort x-ray pulses to specific electronic states and emerging short-lived transient intermediates is of particular relevance for our understanding of fundamental multi-photon absorption processes.
In this work, intense x-ray free-electron laser (XFEL) pulses at the European XFEL (EuXFEL) are combined with a gas cell and grating spectrometer for a high-intensity transmission spectroscopy study of multiphoton-induced ultrafast molecular fragmentation dynamics in CF$_4$.  This approach unlocks the direct intra-pulse observation of transient fragments, including neutral atoms, by their characteristic absorption lines in the transmitted broad-band x-ray spectrum. 
The dynamics with and without initially producing fluorine K-shell holes are studied by tuning the central photon energy. The absorption spectra are measured at different FEL intensities to observe nonlinear effects. 
Transient isolated fluorine atoms and ions are spectroscopically recorded within the ultrashort pulse duration of few tens of femtoseconds. An isosbestic point that signifies the correlated transition between intact neutral CF$_4$ molecules and charged atomic fragments is observed near the fluorine K-edge. The dissociation dynamics and the multiphoton absorption-induced dynamics encoded in the spectra are theoretically interpreted. Overall, this study demonstrates the potential of high-intensity x-ray transmission spectroscopy to study ultrafast molecular dynamics with sensitivity to specific intermediate species and their electronic structure.	
\end{abstract}
                              
\maketitle
\newpage

\section{\label{sec:intro}Introduction}
The interactions of x-rays with matter can create short-lived electronic core-hole states, triggering subsequent ultrafast molecular dynamics~\cite{Gessner2006,Young2010, Hoener2010, Doumy2011,Rudek2012, Travnikova2016, Rudenko2017}. 
The study of such processes is therefore of fundamental importance for diverse research fields such as material science~\cite{Lamberti2016}, photochemistry~\cite{Melanie2003, Chen2005, Sension2020}, biosciences~\cite{Quiney2011, Nass2020} and environmental science~\cite{Lu2015}.
While the linear (low-intensity) absorption and transmission of x-rays in matter is well understood and thus routinely used for scientific applications e.g. at synchrotron light sources, it is desirable to extend our understanding of x-ray transmission in the presence of nonlinear x-ray light-matter interactions. Thus, we ask the basic question, how these interactions manifest in the x-ray transmission spectrum at increasing intensity.

The advent of x-ray free-electron lasers (XFEL) \cite{Emma2010,Seddon2017, Decking2020,McNeil2010,Ueda2018} enabled the nonlinear interaction of more than one x-ray photon with specific atomic sites and their electronic structure~\cite{Young2010, Rudek2012, Doumy2011, Berrah2019, Mazza2020}, opening a new route for site-selective x-ray spectroscopy and control of the molecular dynamics~\cite{Dorfman2016, Rudenko2017, Cavaletto2021}. This is achieved by delivering brilliant x-ray pulses of up to several millijoules focused onto a micrometer-scale focus spot within an ultrashort pulse duration on the order of tens of femtoseconds \cite{Behrens2014, Emma2010, Finetti2017}, and even reaching down to the attosecond regime \cite{Hartmann2018,Duris2020,Li2024,Guo2024}. 

A powerful tool to extract nonlinear dynamical information from transmission spectra is the transient absorption spectroscopy (TAS) approach, which has recently been realized with attosecond pump and probe pulses in the x-ray regime~\cite{Li2024}. Generally, in TAS a broadband pulsed light source is employed to probe the dynamics induced by pump pulses at various time-delays, and spectroscopic information is extracted by resolving the transmitted probe spectra with grating spectrometers.
Conventionally, the probe pulse is weak, therefore, its interaction with the system within TAS is limited to the linear response regime, i.e., measuring the linear absorption spectrum at various delays after a pump pulse, which is typically understood as pump-probe transient absorption spectroscopy.
The absorption spectra however contain additional information when entering the nonlinear response regime of the transmitted pulse, going beyond the pump-probe spectroscopy interpretation.
This has most recently been realized in the XUV regime by using weak HHG-based sources with strong NIR dressing of the coherent dipole response~\cite{Wang2010,Chen2012,Ott2013, Kaldun2016,Stooss2018, Rupprecht2022, Peng2022}, and direct dressing with intense XUV pulses from FEL-based sources~\cite{Ott2019, Ding2019,Ding2021}.
These experiments have revealed for instance modifications of spectral line shapes, which unlock new quantum dynamical information that can be extracted from the absorption imprint of the measured XUV transmission spectra when using intense and ultrashort pulses.
Realizing this approach in the x-ray regime allows for the comprehensive examination of the total absorption of transient states with atomic site- and state-selectivity, including neutral species, simultaneously across a broad range of photon energies and within the ultrashort pulse duration. Hence, the high-intensity x-ray transmission spectroscopy technique developed here can provide a new insight into the ultrafast nonlinear x-ray light-matter interactions even by using just a single pulse and without the need for scanning a time delay.

The CF$_4$ molecule on which we focus below is extensively studied not only because it is a protopypical system for the understanding of ultrafast molecular dynamics~\cite{Ma1991,Bonham1992,Griffiths1993, Glans1994, Saito1994, Itoh1999, Muramats1999, Simone2002, Ueda2003, Guillemin2010, Arion2014, Pertot2017, Iwayama2017}, but also due to its importance in environmental processes, such as the modeling of ozone depletion and climate change~\cite{Lu2015}.
However, to the best of our knowledge, state-resolved fragmentation dynamics of the CF$_4$ molecule nonlinearly driven by intense x-ray fields have not yet been studied. 
In this work, the high-intensity transmission spectroscopy approach is employed to investigate this unexplored area of nonlinear x-ray interaction with a molecular system.

The photon energy of the XFEL short pulses is tuned across the fluorine K-edge to selectively initiate dynamics with and without creating fluorine K-holes after absorbing the first photon at the leading edge of the pulse.
Within high-intensity transmission spectroscopy, the FEL intensity is varied, where we expect the population evolution of different fragments to be influenced by multiphoton absorption.
The spectroscopic fingerprint of these transient fragment states that emerge within the pulse duration is observed via a grating spectrometer.  
In the case where the photon energy lies well below the fluorine K-edge (from around 675 to 680 eV), both the valence and carbon 1s-shell of the CF$_4$ molecule can be ionized, initially forming CF$_4^+$ and C$^*$F$_4^+$ with comparable branching ratio. The cross sections for ionizing carbon 1s and the valence orbitals of CF$_4$ at photon energy 680 eV are 0.12 and 0.117 Mb respectively, approximated by summing the atomic cross section taken from \cite{Vuo}. 
Since the lifetime of the C$^*$F$_4^+$ is only about 8.5 fs~\cite{Nicolas2012}, it will quickly ionize and produce CF$_4^{2+}$ ions via the Auger-Meitner decay process.
We denote this channel as `Auger-before-dissociation' as opposed to another observed process, namely the `dissociation-before-Auger', where the C$^*$F$_4^+$ can dissociate and produce neutral fluorine atoms before Auger-Meitner decay~\cite{Iwayama2017}. 
For the photon energy across the fluorine K-edge, short-lived molecular species CF$^*_4$ and CF$^{*+}_4$ with a fluorine 1s-hole can be respectively created by resonant excitation or photoionization. With a lifetime of around 3 fs~\cite{Arion2014,McCurdy2017}, these molecules will quickly undergo Auger-Meitner decay into predominantly CF$^+_4$ and CF$^{2+}_4$. 
Both molecular ions are unstable, and will thus undergo ultrafast fragmentation processes that produce atomic species~\cite{Pertot2017,Saito1994,Guillemin2010,Ueda2003}. It is these intermediate atomic fragments that are the central focus of our study presented here.
  
The intermediate molecular and atomic fragments of the above initial channels can be observed in the transmitted x-ray absorption spectra across the entire XFEL spectral bandwidth.
To assign the resonant peaks in the spectrum, the electronic structures of relevant intermediate fragments are calculated based on the multiconfiguration self-consistent field (MCSCF) method.     
A semi-classical molecular dynamics (MD) simulation is conducted to rationalize the emergence of atomic fragments within the short pulse duration.
In addition, the sequential multiphoton-absorption-induced dynamics are studied using a simple rate equation model. 
The result of these theoretical calculations allows one to further interpret the measured transmission spectra at high x-ray intensity and thus shed light on the nonlinear physical mechanisms at work in the experiment. 

The paper is organized as follows. 
In Sec.\ref{subsec:assign} we introduce the experimental setup and present measured transmission spectra for photon energies below and across the fluorine K-edge, at various incoming XFEL pulse intensities. 
Precisely calculated resonances of atomic and molecular fragments are presented for the identification of nonlinear multiple ionization and dissociation processes of the CF$_4$ molecule interacting with intense short-pulsed x-ray light. 
The dissociation dynamics of the charged molecules CF$_4^+$ and CF$_4^{2+}$ produced after absorbing the first x-ray photon within the ultrashort pulse is presented in Sec.~\ref{subsec:MD}.
The sequential multiphoton absorption-induced dynamics during the pulse is discussed in Sec.~\ref{subsec:Multiphoton}. 
Finally, a summary and concluding remarks are given in Sec.\ref{sec:con}.

\section{\label{sec:Experiment}Experimental setup and results}
\subsection{\label{subsec:assign}Absorption spectra}
\FloatBarrier
\begin{figure}
	\includegraphics[width=\figurewidth]{{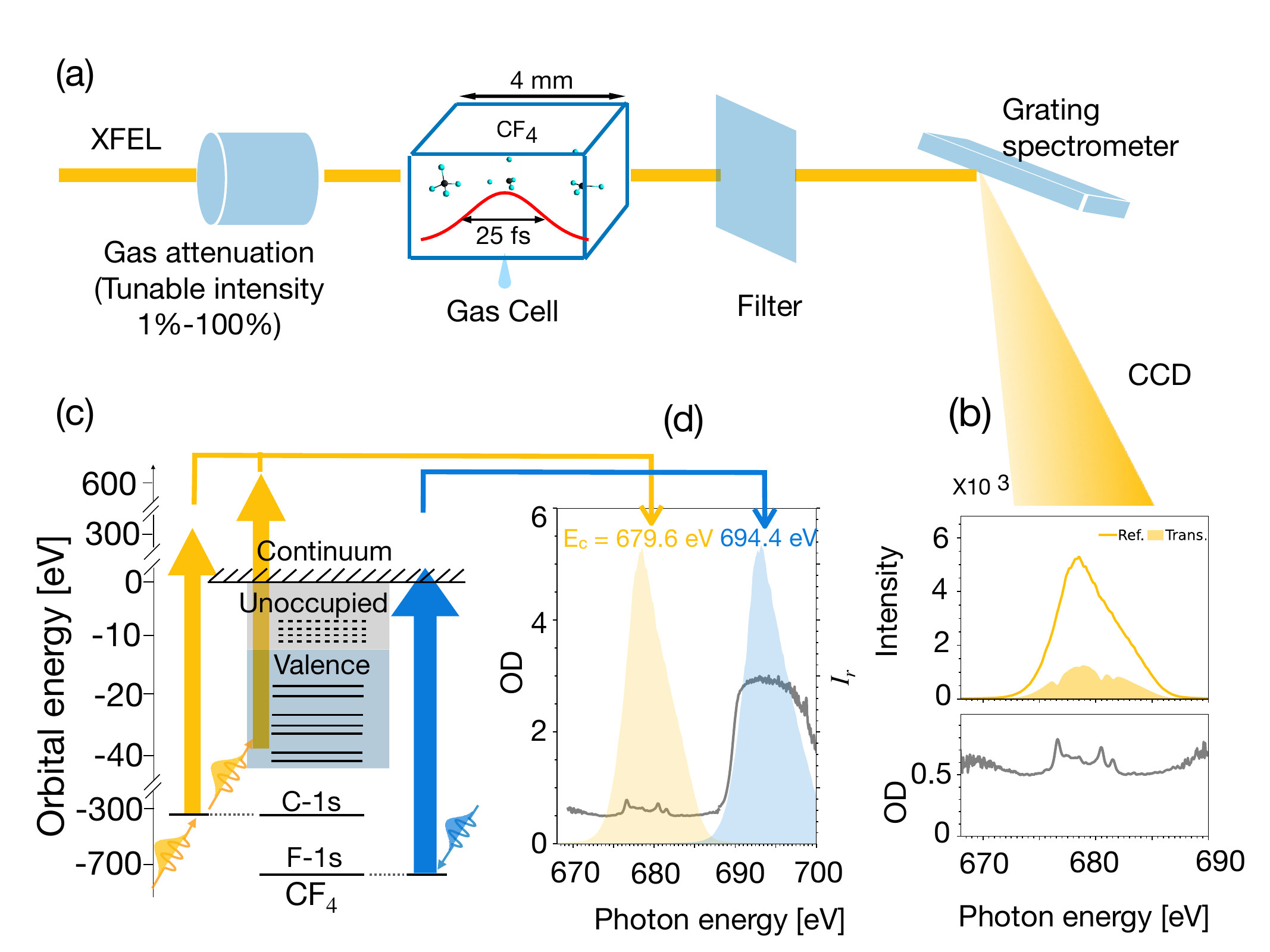}}
	\caption{\label{fig:Experiment}Experimental setup and overview of the x-ray absorption spectra and energy-level scheme for CF${}_4$.(a) Experimental setup. (b) Demonstration of reference and transmitted average spectra of FEL pulses and the derived optical density (OD). (c) Energy diagrams for CF$_4$. The blue and yellow arrows respectively denote the ionization of CF$_4$ with and without creating fluorine K-holes with photon energy centered at 694.4  and 679.6 eV, i.e. above and below the K-edge, respectively. (d) Corresponding OD spectra (gray curve) under the two photon-energy settings. The colored shaded curves show the average spectra of the incoming FEL pulses.}
\end{figure}
The self-amplified spontaneous emission (SASE) free-electron laser EuXFEL is employed in this experiment. 
The experimental scheme is shown in Fig.~\ref{fig:Experiment}(a). 
A short x-ray pulse prepared at the Small Quantum Systems (SQS) instrument with 3 mJ pulse energy and nominal pulse duration of about 25 fs (FWHM) is focused into the CF$_4$ sample in a gas-cell with a 2 $\mu m$ diameter spot size using Kirkpatrick-Baez mirrors. 
In this experiment, the gas-cell pressure is fixed at 0.1 bar and has an inner length of 4 mm in the direction of light propagation.
The FEL pulse travels through a gas attenuator before it is focused into the gas-cell.  
This enables us to measure the photoabsorption spectra at different FEL intensities.   
After passage through the gas cell, the XFEL pulse is sent into a grating spectrometer to facilitate high-intensity x-ray transmission spectroscopy. 
The spectrometer resolution and natural lifetime of the F K-shell state together give an experimental resolution of around 0.2 eV. 
This allows one to separate the resonant structures for electronic states of transient fragments in the total photoabsorption spectra across the entire pulse bandwidth.
The short pulse duration enables observing the intermediate species at the pulse-duration time scales of few tens of femtoseconds.  

\begin{align}\label{eq:OD}
	\mathrm{OD}(E) = -\log_{10}{\frac{\mathcal{I}_{t}(E)}{\mathcal{I}_{r}(E)}} 
\end {align} 
The optical density (OD) is derived employing Eq.~(\ref{eq:OD}), where the shot-averaged intensity spectrum for transmitted $\mathcal{I}_{t}(E)$ and reference $\mathcal{I}_{r}(E)$ beams are separately measured by propagating the x-ray beam through a filled and evacuated gas-cell, respectively, as shown in Fig.~\ref{fig:Experiment}(b).
It is worthwhile to note that, in the high intensity region, Eq.~\ref{eq:OD} does not imply that the conventional single-photon absorption cross section is determined according to the Beer-Lambert's law. 
The energy diagram of the CF$_4$ molecule is presented in (c), which serves to outline the interaction with the first x-ray photon at the beginning of the pulse. 
The ionization potential (IP) for the fluorine 1s orbital in CF$_4$ is about 695.5 eV\cite{Saito1994}.
The central photon energy is set at 679.6 and 694.4 eV to excite and probe different dynamics below and across the fluorine K-edge, as indicated by the yellow and blue arrows, respectively, and color-shaded spectra in the diagram shown in Fig.~\ref{fig:Experiment}(d).

\FloatBarrier
\begin{figure}
	\includegraphics[width=\figurewidth]{{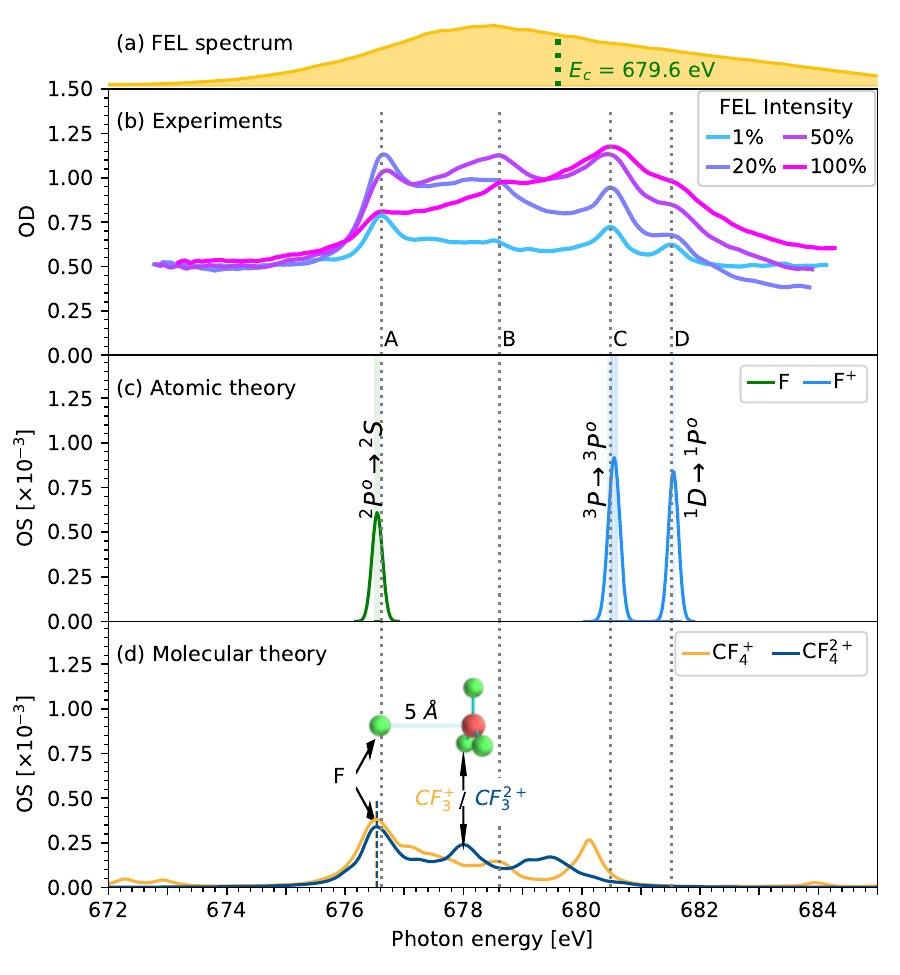}}
	\caption{\label{fig:pre-edge}Resonant absorption spectra for FEL photon energy well below the fluorine K-edge. (a) Average spectrum of incoming FEL pulses centered at 679.6 eV. (b) Experiment OD for various incoming FEL intensities, indicated by the color of the curves. (c) Theoretical calculations for relevant atomic fragments. Note that the calculated oscillator strengths are broadened to match the experimental resolution of about 0.2 eV. (d) Theoretical spectra for fragmenting (dissociating bond length at 5 \AA) charged molecules CF$_4^+$ and CF$_4^{2+}$. The leftmost peaks for these two molecular ions correspond to neutral fluorine fragments which allows to shift the whole molecular spectra to fit peak A. The peaks within 677-679 eV, assigned as CF$_3^+$ and CF$_3^{2+}$, are located in the area of the broad peak B. Each molecular spectrum is calculated by averaging the OS for eight different nuclear geometries displaced from the ground state equilibrium geometry by a combination of C-F dissociation and planarization of CF$_3^{+}$ and CF$_3^{2+}$ (see Appendix~\ref{app:mol-struct}) to describe the vibrational broadening of the peaks, in addition to the 0.2 eV experimental resolution.} 
\end{figure}
The measured OD at different incoming pulse intensities are shown in Fig.~\ref{fig:pre-edge} for central photon energy $E_c=679.6$~eV, well below the fluorine K-edge.  
Four major peaks are observed and marked as A, B, C, and D. 
It is apparent that peak B looks broader than the others, where this difference in width suggests that peak B might be comprised of different transitions e.g. corresponding to molecular fragments while peaks A, C and D are more likely identified as atomic lines.

In order to test this hypothesis, theoretical calculations of the resonant excitation spectra for both atoms and relevant molecular fragments are carried out with high accuracy. 
For the atomic species, the calculation is based on a scheme combining MCSCF and relativistic configuration interaction (RCI)~\cite{Cheng2010} employing the GRASP code~\cite{Fischer2019}.
High calculation accuracy is achieved by a customized quasi-complete basis set and balanced consideration of correlation effects. 
As shown in Appendix~\ref{app:atom-struct}, the transition energies for neutral F atoms, are converged within 0.2~eV. 
In addition, the relative difference between oscillator strengths (OS) in velocity and length gauge is guaranteed within 10\%, indicating the completeness of the basis. 
As for the molecular fragments, the resonant transitions are calculated using the restricted active space self-consistent-field (RASSCF) \cite{Werner1981,Malmqvist1990} and restricted active space perturbation theory (RASPT2) methods \cite{Finley1998,Malmqvist2008} in the OpenMolcas suite\cite{Aquilante2020}.
These methods are reported to provide sub-eV agreement with soft x-ray spectroscopy\cite{Montorsi}. 
The calculation details can be found in Appendix~\ref{app:atom-struct} and \ref{app:mol-struct}.

Two atomic species, F and F$^+$ are found relevant in the spectral region of interest.
Their corresponding OS are indicated as green and blue curves respectively in Fig.~\ref{fig:pre-edge}(c). 
To take into account the experimental resolution, the theoretical spectra were broadened (by convolution) to 0.2 eV. 
The calculated resonances of F (1s$^2$2s$^2$2p$^5$ $\to$ 1s$^1$2s$^2$2p$^6$) and F$^+$ (1s$^2$2s$^2$2p$^4$ $\to$ 1s$^1$2s$^2$2p$^5$) are in good agreement with the experimental peaks A, C and D.
Note that the two electronic states 1s$^2$2s$^2$2p$^4\ {}^3$P and ${}^1$D for the F$^+$ initial state are individually resolved. 
Assigning the molecular fragments with their complex and closely-spaced electronic and vibrational states is more challenging, in addition to the quantum molecular calculation itself. 
Therefore, we focus on the F-1s resonant transition of dissociating charged molecules after absorbing the first photon, as depicted by the inset cartoon in subfigure (d).
Both singly charged CF$_4^+$ or doubly charged CF$_4^{2+}$ (with the dissociating C-F bond distance set at 5 \AA) are calculated and shown as yellow and dark blue curves in (d).
The molecular spectra are broadened by vibrational coupling to nuclear motion in addition to the experimental resolution of 0.2 eV, described by averaging the OS of eight different nuclear
geometries displaced by a combination of C-F dissociation and planarization of CF$_3^+$ and CF$_3^{2+}$~\cite{Kurt1997, Li2024, Takahashi2019} (see Appendix \ref{app:mol-struct}).
The wavefunction analysis shows the first peak corresponds to a neutral fluorine atom in both cases, this allows us to shift the whole theoretical molecular spectra to fit the first experimental peak A, which is assigned to a neutral fluorine atom separately by the highly accurate atomic calculation.
Stark shift due to the nearby charged radicals CF$_3^+$ and CF$_3^{2+}$ are taken into account respectively, indicated by the dashed lines slightly offset from peak A (-0.09 eV for CF$_3^+$ and -0.08 eV for CF$_3^{2+}$ at a distance of 5 $\AA$ to the dissociating F atom used in our calculations).
As a result, the transition OS from both radicals CF$_3^+$ and  CF$_3^{2+}$ falls into the range of the broad peak B. The agreement of the theoretical prediction with the experimental spectra thus confirms the above hypothesis of assigning the narrower peaks A, C, and D predominantly to neutral F and respectively F$^+$ in different initial states as indicated in Fig.~\ref{fig:pre-edge}(c), as well as the wider peak B to be influenced by absorption of molecular fragments.

\FloatBarrier
\begin{figure}
	\includegraphics[width=\figurewidth]{{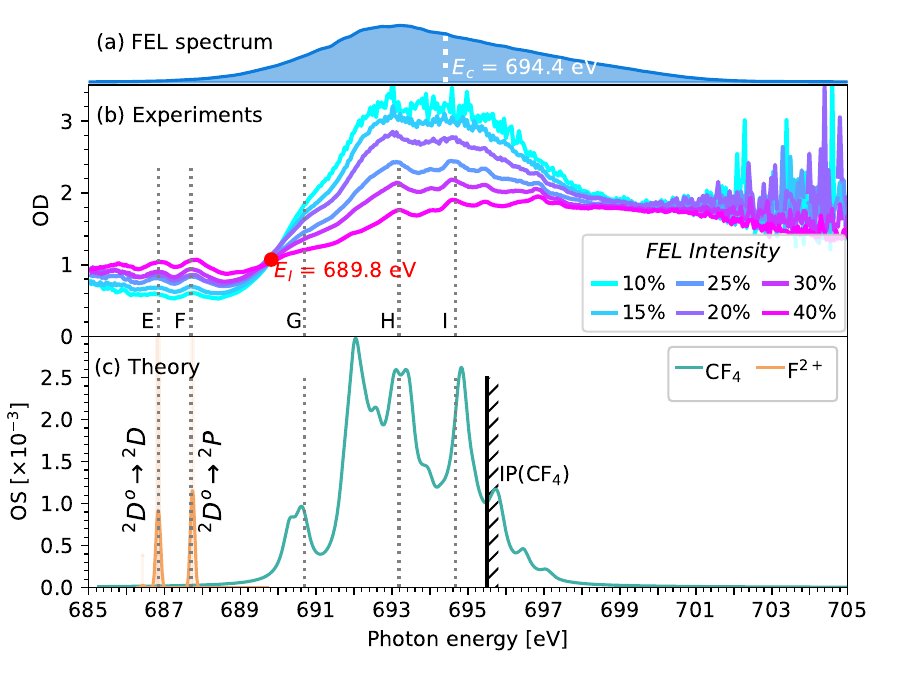}}
	\caption{\label{fig:edge} Resonant absorption spectra across the fluorine K-edge. (a) Average spectrum of incoming FEL pulses with central photon energy at 694.4 eV. (b) Experiment OD for various incoming FEL intensities. (c) Theoretical calculations for relevant atomic and molecular fragments. The molecular spectrum is broadened by vibrational coupling to the combination of C-F dissociation and planarization of CF$_3^+$  and CF$_3^{2+}$ in addition to the 0.2 eV experiment resolution. The ionization potential (IP) for CF$_4$ is indicated by the vertical shaded bar. }
\end{figure}
To study the nonlinear dynamics initiated by F-1s holes, the central photon energy was shifted to 694.4 eV.  
The spectra for different FEL intensities are shown as colored curves in Fig.~\ref{fig:edge}(b).
In this region, the general shape of a large and broad peak on top of the F-1s edge qualitatively matches with the linear synchrotron studies~\cite{Simone2002,Arion2014}.
Still, additional detailed resonant structure is visible, even for the lowest FEL intensity used (10\%), where we label five obvious resonant peaks from E to I for later discussion.
Atomic structure calculations are used to assign peaks E and F to doubly charged atoms F$^{2+}$ in its ground states (orange curve in Fig.~\ref{fig:edge}c), produced within the pulse upon nonlinear interaction, which is also supported by the fact that they are absent in linear synchrotron measurements~\cite{Simone2002}. 
The resonant transition peak of CF$_4$ lies at the center of the big hump across the edge.
Three peaks G, H and I can be identified in all curves, they correspond to the three antibonding transition final-states, which are also observed in the linear synchrotron absorption spectrum\cite{Simone2002}. 
The theoretical spectrum was shifted by -1.0 eV to align it with the reference experimental synchrotron spectra~\cite{Simone2002}. 
The theoretical spectrum was calculated by averaging the OS of eight different nuclear geometries displaced by the umbrella motion which is the mostly IR-active motion at the equilibrium geometry of CF$_4$ (see Appendix~\ref{app:mol-struct}), to describe the vibrational broadening of the peaks, in addition to the experimental resolution.
The overall broadened spectra manifest features that agree well with the three peaks fitted in the synchrotron spectra~\cite{Simone2002}.
Moreover, we observe additional sharp peaks emerge at higher intensities, hypothetically due to molecular two-sited double core-hole states or higher charged atomic ions, as well as due to non-thermal motion competing with the molecular geometric distortion of the excited neutral molecule after the absorption of the first x-ray photon.
Given the high number of possible states, further experimental and theoretical investigation is needed to further clarify their origin, which goes far beyond the scope of this work.
  
A pronounced isosbestic point\cite{Braslavsky2007} at $E_i=$689.8 eV is observed and marked as a red dot in Fig.~\ref{fig:edge}a. 
To be specific, the OD for energy $E<E_i$ increases with the increasing FEL pulse intensity, while the OD for energy $E>E_i$ decreases with increasing FEL intensity.
Due to strong photoabsorption processes in the energy region of interest, the density of available neutral CF$_4$ molecules within the FEL pulse duration drops with increasing FEL intensity, leading to the lowering of its absorption signal above the isosbestic point.
Correspondingly, the absorption signal in the region of peaks E and F increases due to the increasing density of the F$^{2+}$ fragments.

\subsection{\label{subsec:MD}Single-photon-induced molecular dynamics}

\begin{figure}[h!]
	\includegraphics[width=\figurewidth]{{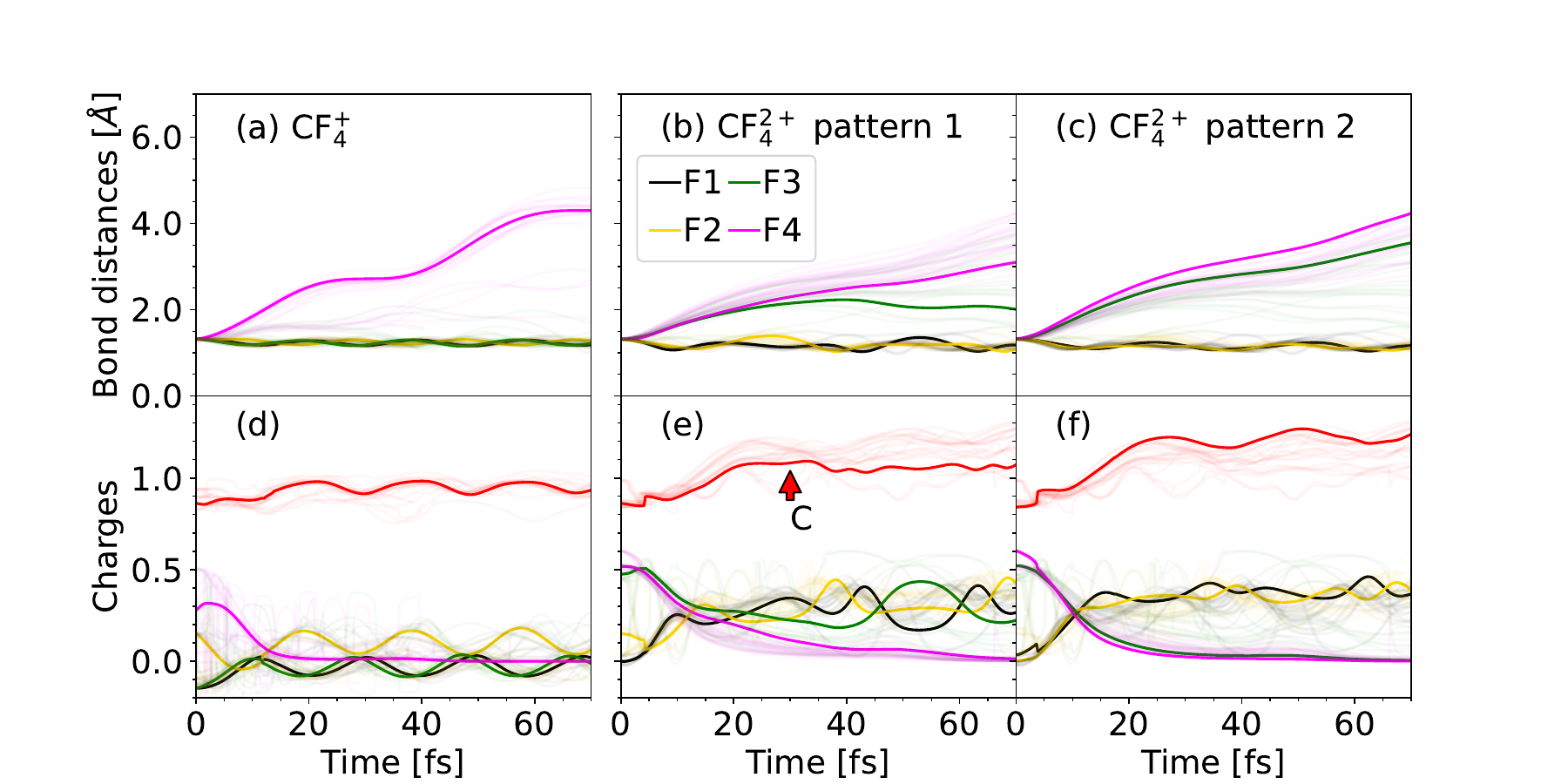}}
	\caption{\label{fig:RASMD} Molecular fragmentation dynamics for both CF$_4^+$ (created by valence ionization) and CF$_4^{2+}$ (created by photonionizing the C-1s hole and following Auger-Meitner decay) before the arrival of a second photon. (a), (b) and (c) Bond distances for CF$_4^+$ and CF$_4^{2+}$ respectively. The molecular fragmentation of CF$_4^{2+}$ starting from 1t$_1^{-2}$ and 4t$_2^{-1}$1t$_1^{-1}$ are highlighted in (b) and (c) respectively. (d), (e) and (f) correspond to charge on each atomic site. Note that the charge on the distant F atom (F4) vanishes within few tens of femtosecond time scale for all cases, indicating the atomic fragments are neutral. }
\end{figure}
In the previous section, we have assigned atomic fragments (neutral and charged) to several peaks. However, it is necessary to quantitatively study the dissociation dynamics of the products after absorbing the first x-ray photon within the leading edge of the pulse, as a foundation for the later study of multiphoton absorption dynamics in Sec.~\ref{subsec:Multiphoton}. 
Photons with energy around 680 eV can either ionize electrons out of the valence shell of the CF$_4$ molecule, or from the localized carbon 1s core orbital, both with comparable cross-sections. 
While the former channel produces mainly CF$_4^{+}$ ions, the latter channel can quickly render predominantly doubly charged CF$_4^{2+}$ ions after Auger-Meitner decay within 8.5~fs\cite{Nicolas2012}. The double Auger decay (creating triply charged ion CF$_4^{3+}$) is neglected due to its reported low possibility of around 10\%~\cite{Roos2018}. 
To verify the possibility of subsequent ultrafast fragmentation within the short pulse duration, we carried out a semiclassical molecular dynamics (MD) simulation using the DYNAMIX module in OpenMolcas, which is based on state-averaged CASSCF energy surfaces.
To study the dissociation dynamics for different ionization products with various hole distribution possibilities, we carry out six separate sets of MD simulations where molecules are initially put on different energy surfaces. Each energy surface represents an electronic state with one hole in the six highest-occupied valence orbitals for CF$_4^+$, and two holes in the four highest-occupied valence orbitals for CF$_4^{2+}$.
Surface hopping is used to account for possible nonadiabatic transitions.
As a result, the molecules with highly valence-excited states may end up in lower electronic states when passing the conical intersections, and give off excessive energy to the nuclear kinetic energies, thus enabling or accelerating the dissociation process (for a demonstration example, see Fig.~\ref{fig:hop} in Appendix~\ref{app:ras-md}). 
Each set consists of an ensemble of five parallel simulations with various initial nuclear velocities to take into account a finite temperature of 300 K.    
Further simulation details can be found in Appendix~\ref{app:ras-md}. 

The MD simulation results of CF$_4^+$ and CF$_4^{2+}$ are shown in Fig.\ref{fig:RASMD}. 
The simulation for CF$_4^+$ from all six initial states and with various initial velocities show similar fragmentation pattern, as depicted by the transparent curves in the first column. The fragmentation dynamics from the ground state 1t$_1^{-1}$ is highlighted as a representative case.
Subfigure (a) shows the bond distances of four fluorine atoms (denoted as F1, F2, F3, and F4) to the carbon center. 
One of the fluorine atoms (F4) quickly flies away from the carbon center, leaving the remaining fluorine atoms vibrating around their initial bond distances. 
In subfigure (d) we observe this dissociating F4 atom to carry away around 0.3 charge initially, however, due to its strong electronegativity, its charge quickly drops to zero after around 20~fs.
We take this time scale of an ejected neutral fluorine atom as approximated dissociation time $\tau_D$ of CF$_4^+$ for later discussion in Sec.~\ref{subsec:Multiphoton}.
As for the CF$_4^{2+}$ ion, the dissociation dynamics can be classified into two typical trends, and highlighted in the second and third columns of Fig.~\ref{fig:RASMD}. 
In subfigures (b) and (e), the molecular dynamics starting from the ground state 1t$_1^{-2}$ is highlighted to represent a typical case, where one fluorine atom flies away, with its charge eventually turning neutral within a time scale of about 60 fs.
For the molecular dynamics starting from the excited state 4t$_2^{-1}$1t$_1^{-1}$, as highlighted in subfigures (c) and (f), we observe two fluorine atoms to be ejected, where both of them turn neutral within about 40 fs.
This study therefore confirms that after the interaction with the first x-ray photon at the leading edge of the pulse, the charged molecule can indeed fragment within few tens of femtosecond and thus on the time scale or shorter than the pulse duration. This further supports the above identification of neutral fluorine atoms spectroscopically observed within the same x-ray pulse. 

\subsection{\label{subsec:Multiphoton}Multiphoton dynamics}

\begin{figure}
	\includegraphics[width=\figurewidth]{{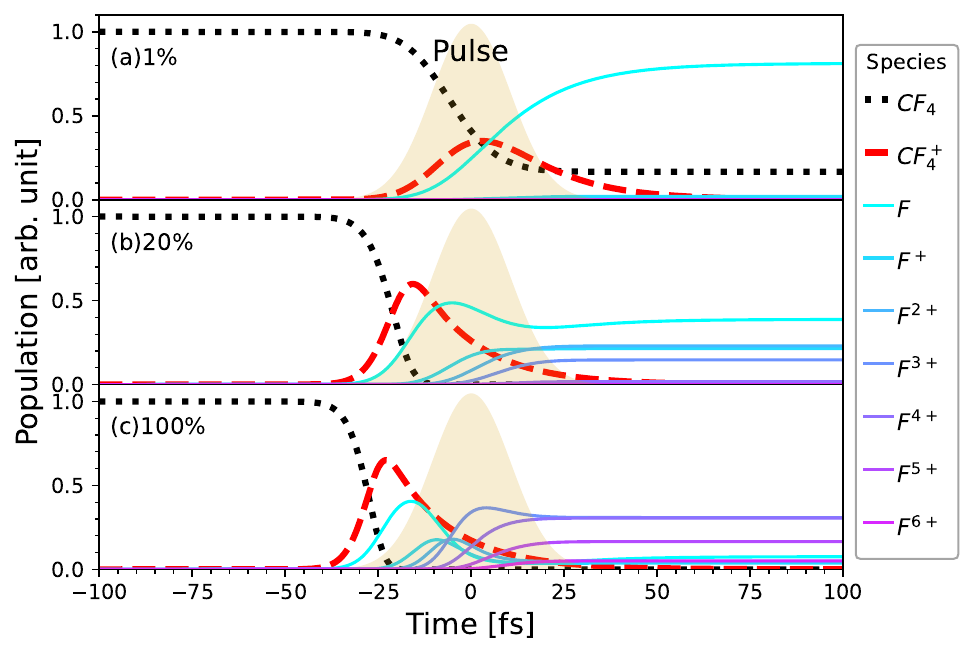}}
	\caption{\label{fig:temp-pop} Evolution of fragment population for (a) 1\% FEL intensity, (b) 20\% FEL intensity and (c) 100\% FEL intensity. The FEL pulse profile used in the model is shown as a filled shaded curve.}
\end{figure}

\FloatBarrier
Having studied the dissociation dynamics after absorbing the first photon, we now turn to the sequential multiphoton absorption processes within the FEL pulse to discuss their nonlinear intensity dependence, regarding atomic peaks A, C, and D well below the fluorine K edge.
A simplified rate equation model is employed to study the sequential absorption of further x-ray photons following the molecular dissociation.
For simplicity, the appearance of neutral fluorine in various fragmentation patterns studied in Sec.~\ref{subsec:MD} is phenomenologically described by an effective channel. In this channel, the CF$_4$ is photoionized (regardless of whether it's valence or C-1s ionization) into CF$_4^+$, which dissociates and effectively produces a neutral F atom in the time scale of $\tau_D$ of 20 fs, as determined from the MD simulation in Sec.\ref{subsec:MD}.
This approximation is validated for the current focus of study, i.e., the atomic species shown as peaks A, C and D in Fig.~\ref{fig:pre-edge}b, with a trade-off for ignoring the details of partner molecular fragments assigned in the area of peak B.
The possible direct creation of singly charged fluorine F$^+$ from much higher electronic states of CF$_4^+$~\cite{Lapiano‐Smith1989} is ignored.
In addition, the `dissociation-before-Auger' channel is also ignored due to the challenge of detailing the molecular fragmentation with core-hole.
Within these approximations, the charged fluorine atoms are then only produced by sequentially ionizing these neutral F atoms to higher charge states.
The considered atomic x-ray light-matter interaction processes are photoionization, resonant excitation, and subsequent decay by fluorescence or Auger-Meitner decay. The corresponding reaction rates $\Gamma_{ij}$ for the transition from electronic configuration $i$ to $j$ are obtained by using the configuration interaction method via flexible atomic code (FAC)\cite{Gu2004}. For the upward transitions ($i<j$), $\Gamma_{ij} = \sigma^{PI}_{ij}J(t) + \sigma^{RE}_{ij}J(t) + A_{ij}$, where $\sigma^{PI}_{ij}$, $\sigma^{RE}_{ij}$ and $A_{ij}$ denote the cross-sections of photoionization, resonant photoexcitation and the rate of Auger-Meitner decay respectively.
Rate matrix elements for the downward transitions ($i>j$) is simply the rate of fluorescence decay $\Gamma_{ij} = R_{ij}$.
The set of rate equations for the populations $N_X$ reads,
\begin{align}
	&\frac{dN_{\scriptscriptstyle CF_4}}{dt} = - \left[\sigma_{\scriptscriptstyle CF_4}J(t)\right] N_{_{\scriptscriptstyle CF_4}}, \tag{2-a}\label{eq:REQa}\\
    &\frac{dN_{\scriptscriptstyle CF_4^+}}{dt} =  \left[\sigma_{\scriptscriptstyle CF_4}J(t)\right] N_{\scriptscriptstyle CF_4} -\frac{1}{\tau_{\scriptstyle D}}N_{\scriptscriptstyle CF_4^+} \tag{2-b}\label{eq:REQb}\\
	&\frac{dN_0}{dt} = \sum_{j\ne 0}N_j\Gamma_{j0}-N_0\sum_{j\ne 0}\Gamma_{0j} + \frac{1}{\tau_{\scriptstyle D} }N_{\scriptscriptstyle CF_4^+} \tag{2-c}\label{eq:REQc}\\
	&...\notag \\
    &\frac{dN_i}{dt} = \sum_{j\ne i}N_j\Gamma_{ji}-N_i\sum_{j\ne i}\Gamma_{ij}, \tag{2-d}\label{eq:REQd}
\end{align}
where the index $X$ denotes both the neutral CF$_4$ and singly ionized CF$_4^{+}$ molecular species, as well as the fluorine atom in configuration $i$ ($0\le i\le 60$ indexes all configurations ranging from neutral F to F$^{8+}$). 
Eq.~(\ref{eq:REQa}) describes the molecular ionization and Eq.~(\ref{eq:REQb}) describes the evolution of CF$_4^+$ due to the competition of ionization of CF$_4$ and following dissociation of CF$_4^+$.
The emergence of neutral F atom due to the dissociation of CF$_4^+$ corresponds to the source term of $\frac{1}{\tau_D}N_{CF_4^+}$ in Eq.~(\ref{eq:REQc}).
The FEL photon flux $J(t)$ is modeled with a Gaussian temporal pulse profile with a pulse duration of 25 fs (FWHM) and a center photon energy of 676 eV. The molecular photoionization cross section $\sigma_{\scriptscriptstyle CF_4}=0.24$~Mbarn is approximated by the summation of the cross sections of four fluorine atoms and a carbon atom taken from the VUO database~\cite{Vuo}.
The resulting population dynamics for the CF$_4$, CF$_4^+$ as well the atomic species at different FEL photon flux $J(t)$ are shown in Fig.~\ref{fig:temp-pop}.
The populations of the atomic species shown here are charge state resolved, i.e., the population of the configurations with the same charge are summed up as $P_Q = \sum_{i=0}^{60}N_i\delta_{Q_i,Q}$ ($Q_i$ is the charge of the configuration $i$). 
In the low FEL intensity case in Fig.~\ref{fig:temp-pop}(a), the neutral fluorine population (blue curve) exponentially rises within the pulse as a result of single-photon induced fragmentation and subsequent dissociation. This is in agreement with the decrease of the initial CF$_4$ molecule (black dotted curve) as well as the transient rise and decay of the intermediate CF$_4^+$ molecular cation (red dashed curve).
However, as the FEL intensity further increases in subfigures (b) and (c), saturation sets in, where the neutral fluorine population starts to form a peak, and eventually decreases in favor of higher charge states. 
For high intensities, the neutral fluorine is nearly entirely bleached at the end of the pulse in (c), but the transient appearance of this channel during the FEL pulse can still be recorded in the measured absorption spectra, identified as peak A in the magenta curve in Fig.~\ref{fig:pre-edge}(b).
This demonstrates a key advantage of the high-intensity x-ray transmission spectroscopy scheme, being sensitive to intra-pulse intermediate states.
\FloatBarrier

\FloatBarrier
\begin{figure}
	\includegraphics[width=\figurewidth]{{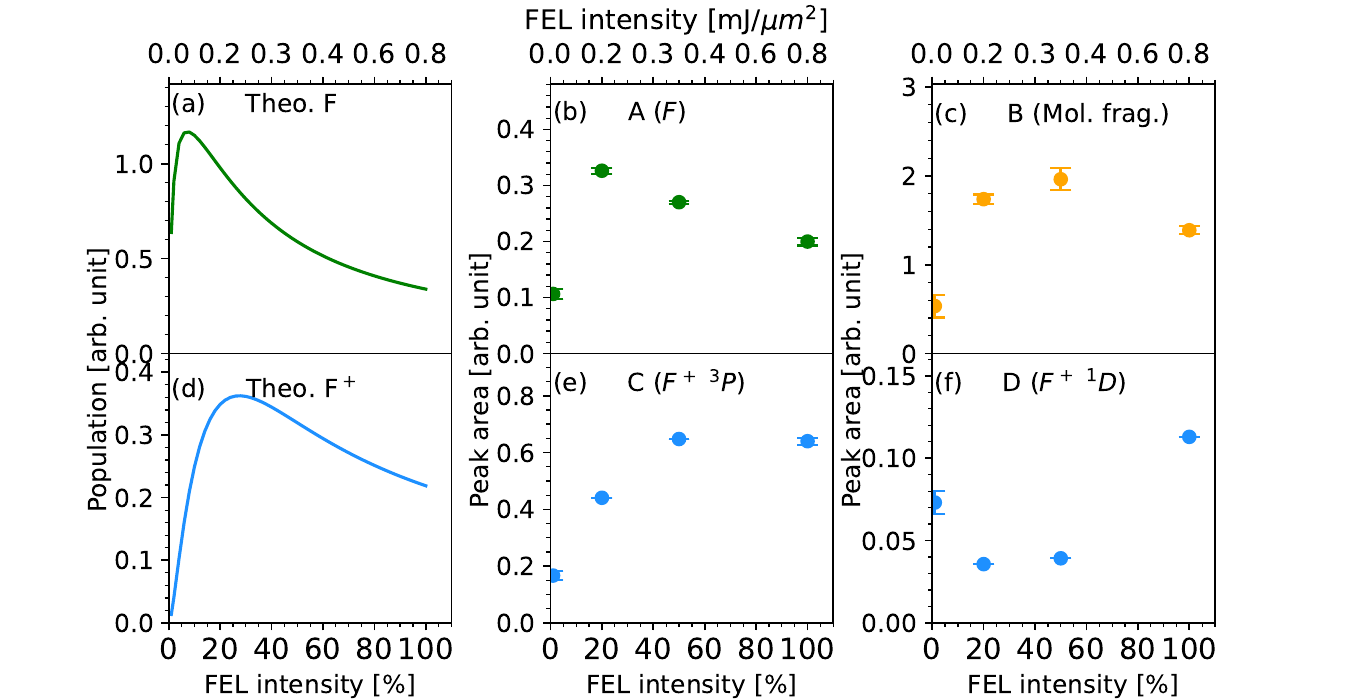}}
	\caption{\label{fig:multiph-pop} Pulse integrated populations in comparison with the fitted peak areas. 
     (a) and (d): The pulse integrated populations of neutral and singly charged fluorine atoms as functions of FEL intensity. 
     (b), (c) (e) and (f): the fitted area of the peaks A, B, C, and D in Fig.~\ref{fig:pre-edge}.}
\end{figure}
A direct comparison of the rate-equation model with the experiment can be done by temporally integrating the resulting population dynamics with the normalized FEL pulse profile. The result is shown in Fig.~\ref{fig:multiph-pop}, where the pulse-integrated populations of the neutral and singly ionized atomic products (F and F$^+$) are shown in comparison with the fitted peak areas of the experimental spectra from Fig.~\ref{fig:pre-edge}.
The population of neutral fluorine reaches a local maximum at around 10\% FEL intensity, which is in agreement with the formation of a local maximum of peak A at relatively low FEL intensity shown in (b).

The simulated population of singly charged fluorine in subfigure (d) reaches its local maximum only at higher FEL intensity, where a similar trend is observed in peak C in subfigure (e), which quantifies the appearance of F$^+$ in the $^3$P initial state. It is interesting to note that the appearance of F$^+$ in the $^1$D initial state shows a different trend of reaching a local minimum at intermediate intensity, however at a much lower absolute scale of the peak area. This trend is currently unexplained in the simple rate-equation model, however, we also cannot rule out the contribution of other unidentified fragments in the spectral region of peak D at high FEL intensity, especially since we notice the absorption peak to significantly broaden in Fig.~\ref{fig:pre-edge}(b). Nevertheless, the different trends of peaks C and D highlight the possibility to discern transient species with sensitivity to their electronic state. 
Given the simplicity of the rate-equation model, the overall agreement with the experiment is remarkable.

\section{\label{sec:con}Conclusion}

In summary, we used intense ultrashort x-ray FEL pulses in a high-intensity transmission spectroscopy  to study multi-photon induced molecular dynamics in CF$_4$ molecules. The combination of a gas-cell and grating spectrometer to analyze the FEL spectrum after transmission through the gas-phase molecular sample allows us to directly observe the transient intermediate fragments, including the neutral ones, within the pulse duration of few tens of femtoseconds.  The energy resolution of around 0.2 eV allows for the identification of different electronic states.  Measuring absorption spectra at different FEL pulse intensities allows to change the abundance of different fragment species, resolved in charge and electronic states.

Dynamics with and without initially producing fluorine K-shell holes at the leading edge of the pulse have been investigated with the FEL central photon energy tuned to 694.4 and 679.6 eV, respectively. 
In the spectra well below the F-K edge, neutral fluorine atoms are spectroscopically observed within the nominal 25 fs (FWHM) pulse duration. 
Along with transiently appearing neutral atoms, singly charged fluorine atoms as well as CF$_3^+$ and CF$_3^{2+}$ molecular fragments are spectroscopically identified with the help of meticulous atomic and molecular electronic structure calculations. 
In the measurement across the F-K edge, absorption spectra distinct from synchrotron measurements are observed at different FEL intensities. Five major peaks in the spectra are identified, including F$^{2+}$ atoms, along with an isosbestic point at 689.8~eV that signifies the correlated transition between intact neutral CF$_4$ molecules and charged atomic fragments.

A semi-classical molecular dynamics simulation of the initial fragmentation dynamics of the charged molecules after the absorption of the first photon at the leading edge of the pulse has been carried out to understand the ultrafast dissociation into neutral fluorine atoms.
Based on this, a system of rate equations is used to model the transient intermediate population dynamics. The pulse-integrated population of fluorine atoms and ions at different FEL intensities has been obtained and agrees with the spectroscopically identified peak areas in the experiment. 
     
In the future, such experiments can provide new benchmark data for the ongoing development of more sophisticated time-resolved molecular spectroscopy calculations for the interaction of matter with intense ultrashort x-ray pulses.
On the experimental side, these initial results motivate to further develop new schemes for time-resolved x-ray pump x-ray probe transient absorption spectroscopy with a two-pulse setup at few and sub-femtosecond time resolution~\cite{Li2024,Guo2024,Rebholz2021}. These developments will eventually help disentangle the competing pathways of electronic and structural dynamics in ultrafast nonlinear x-ray light-matter interactions.

Data recorded for the experiment at the European XFEL are available at~\cite{DATA}.
 
\begin{acknowledgments}
We acknowledge European XFEL in Schenefeld, Germany for provision of the x-ray free-electron laser-beam time at the SQS instrument and would like to thank the staff for their assistance. We acknowledge Christian Kaiser and Alexander von der Dellen and the technical workshop of MPIK for their support. We thank Nina Rohringer for the provision of the x-ray CCD detector.
RJ would like to thank Alexander Kuleff for insightful discussions.
AF, YN, KL, GD, and LY were supported by the US Department of Energy, Office of Science, Basic Energy Sciences, Chemical Sciences, Geosciences, and Biosciences Division under award DEAC02-06CH11357.
\end{acknowledgments}

\appendix
\section{Calculation of atomic structures}\label{app:atom-struct}
For a complete survey of the possible atomic fragments that appeared in the experimental spectra, multiple types of atomic core-excitation transition patterns need to be considered, depending on the initial and final states of the transitions. 
Take the neutral F for example, the initial states of the transition can also be in low-lying valence excited configuration, e.g., 1s$^2$2s$^2$2p$^4$3s$\to$ 1s$^1$2s$^2$2p$^5$3s, where there is one standing-by electron in 3s orbital. 
In addition, the 1s electron can be excited to either 2p orbital or Rydberg orbitals (i.e., 3p, 4p, et al.), i.e., 1s$^2$2s$^2$2p$^5 \to$1s$^1$2s$^2$2p$^6$ or 1s$^1$2s$^2$2p$^5$3p.
Therefore, multiple electronic states must be taken into account for both initial and final states.
In this work, the calculation of resonant core-excitation transitions for fluorine atoms (ions) is based on a combined multi-configuration self-consistent field (MCSCF) and relativistic configuration interaction (RCI) scheme\cite{Cheng2010} employing the GRASP code\cite{Fischer2019}. The general working flow is as follows: the atomic basis sets for the initial and final states of the transitions are optimized separately using the MCSCF method. Then individual RCI calculations of initial and final states are carried out based on the corresponding atomic basis respectively. 
Since the atomic basis for initial and final states are optimized individually, they are nonorthogonal to each other, therefore we need to transform the two basis sets (and CI coefficients accordingly) to make them biorthonormal~\cite{Olsen1995} using the rbiotransform module of the GRASP code. 
Then the transition oscillator strengths between the initial and final states are calculated using the rtransition module of GRASP code\cite{Olsen1995,Fischer2019}.

The atomic orbital basis is optimized layer by layer, meaning that, we start from optimizing a minimum basis set using MCSCF, then we extend the basis set in a new MCSCF calculation, but keep the previously optimized orbitals fixed. We repeat this procedure until the convergence is achieved.
Taking the neutral F atoms as an illustration example, the orbitals are optimized in 7 layers: \{1s,2s,2p\}, \{3s\}, \{3p\}, \{3d,4$l$\}, \{$5l$\},\{$6l$\}, \{$7l$\}, ($l\le\min{\{n-1,4\}}$).
The first layer of \{1s,2s,2p\} orbitals are obtained by optimizing the $~^2$P$_{  3/2,1/2  }^o$ states (using configuration 1s$^2$2s$^2$2p$^5$). 
For the second layer, the \{3s\} orbital is obtained by optimizing the $^4$P$_{  5/2,3/2,1/2}$, $^2$P$_{  3/2,1/2}$, and $^2$D$_{  5/2  }$ states (using configuration 1s$^2$2s$^2$2p$^4$3s) with the first layer fixed.
For the third layer, the \{3p\} orbital are obtained by optimizing the $^4$P$_{  5/2,3/2,1/2  }^o$, $^4$D$_{  5/2  }^o$ states (using configuration 1s$^2$2s$^2$2p$^4$3p), with the first and second layers fixed. 
Note that these states will be used to calculate transition rates, therefore the radial wavefunctions of 1s to 3p are treated as spectral orbitals (nodal structure guaranteed during orbital optimization).
The fourth layer \{3d,$4l$\} is obtained by adding 3d and $4l$ orbitals in addition to all the above optimized orbitals \{1s,2s,2p\}, \{3s\}, \{3p\}, and optimizing all the above 13 energy states together, based on the configuration state functions (CSF) generated by allowing single, double, and some important triple electron permutations from occupied orbitals in reference CSFs (i.e.,1s, 2s, 2p, 3s, 3p) to the \{3d, $4l$\} ($l<4$).
We repeat this procedure for the \{$5l$\}, \{$6l$\}, \{$7l$\} ($l\le4$) successively to further include the dynamical correlations.
From \{3d\} onwards, the orbitals are treated as so-called pseudo orbitals instead, i.e., their nodal structure is not guaranteed during optimization.
The orbitals for the final states of the core-excitation transition can be obtained in the same manner, except for the additional requirement that at least one K-hole should be present in all CSFs.

Then all the energy levels of interest can be obtained with the optimized orbital basis set. The energy levels of the 13 relevant initial states calculated in different correlation models are listed in Tab.~\ref{tab:dE}. Correlation model $3sp$ corresponds to the basis with the first 3 layers, $n4$ corresponds to the first 4 layers of the basis, and so on. The best calculation $n7$ shows less than 1\% discrepancy in comparison with the NIST database~\cite{NIST}.
The energy of the 1s$\to$2p transitions calculated from different correlation models are listed in Tab.~\ref{tab:Etrans}, which is converging to the experimental fit value within 0.2~eV. The relative difference between oscillator strength (OS) in velocity (V-) and length (L-) gauge representation is guaranteed within 10\%, indicating the completeness of the computation basis. 
\FloatBarrier
\begin{table}
	\centering
	\caption{\label{tab:dE}The energies (in eV) of the initial states (including excited configuration) of core-excitation transition for neutral F, calculated with different levels of electronic correlation. NIST\cite{NIST} atomic levels are listed for comparison.} 
    \footnotesize
   	\begin{ruledtabular} 
	 \begin{tabular}{c@{}cc@{ }c@{ }c@{}c@{}c@{}c@{}c@{}c@{}c}
		\centering
	{} & 	Term  &   $3sp$   &    $n4$  &    $n5$   &    $n6$   & $n7$      & NIST \\
			\hline
\parbox{2cm}{Ground state}    &  2p$^5~^2$P$_{  3/2  }^o$ &    -2707.10   &  -2712.82    &   -2714.42   &  -2714.80    & -2715.02   & --  \\
			\hline
       \multirow{12}{*}{\centering \parbox{2cm}{Energy relative to ground state 2p$^5~^2$P$_{  3/2  }^o$ }} &     
         2p$^5~^2$P$_{  1/2  }^o$ &   0.0483   &   0.0495 &    0.0502 &    0.0495 &   0.0498   & 0.0501  \\  
        &2p$^4$3s$~^4$P$_{  5/2  }$    &   13.14 & 12.75 & 12.94  & 12.75 & 12.73  & 12.70  \\
        &2p$^4$3s$~^4$P$_{  3/2  }$    &   13.17 & 12.79 & 12.98  & 12.79 & 12.76  & 12.73  \\  
        &2p$^4$3s$~^4$P$_{  1/2  }$    &   13.19 & 12.81 & 13.00  & 12.81 & 12.78  & 12.75  \\ 
        &2p$^4$3s$~^2$P$_{  3/2  }$    &   13.51 & 13.11 & 13.28  & 13.11 & 13.06  & 12.98  \\ 
        &2p$^4$3s$~^2$P$_{  1/2  }$    &   13.55 & 13.15 & 13.32  & 13.15 & 13.10  & 13.03  \\ 
        &2p$^4$3p$~^4$P$_{  5/2  }^o$  &   14.82 & 14.43 & 14.52  & 14.43 & 14.40  & 14.37  \\  
        &2p$^4$3p$~^4$P$_{  3/2  }^o$  &   14.83 & 14.44 & 14.53  & 14.44 & 14.41  & 14.39  \\ 
        &2p$^4$3p$~^4$P$_{  1/2  }^o$  &   14.84 & 14.45 & 14.54  & 14.45 & 14.42  & 14.40  \\  
        &2p$^4$3p$~^4$D$_{  5/2  }^o$  &   15.03 & 14.62 & 14.71  & 14.62 & 14.58  & 14.53  \\ 
        &2p$^4$3s$~^2$D$_{  5/2,3/2  }$    &   15.85 & 15.47 & 15.62  & 15.47 & 15.42  & 15.36 \\  
\end{tabular}
\end{ruledtabular}
\end{table}

\begin{table}
	\centering
	\caption{\label{tab:Etrans}The transition energies (in eV) for the neutral F 1s$\to$2p, calculated with different levels of correlations. The experimental fit values are listed for comparison.} 
    \footnotesize
   	\begin{ruledtabular}     
		\begin{tabular}{r@{}cc@{ }c@{ }c@{}c@{}c@{}c@{}c@{}c@{}c} 
		\centering
	     Initial states  &   $3sp$   &    $n4$  &    $n5$   &    $n6$   & $n7$      & Exp. \\
	     \hline
          $~^2P_{  1/2  }^o$  & 677.58	&675.90	&676.38	&676.40	&676.51  & \multirow[c]{2}{*}{676.60} \\    
          $~^2P_{  3/2  }^o$  & 677.63	&675.95	&676.43	&676.45	&676.56  &   \\
		\end{tabular}
\end{ruledtabular}	
\end{table}

\normalsize

\section{Calculation of molecular structures}\label{app:mol-struct}

The electronic structure of three relevant molecular species CF$_4$, CF$_4^{+}$, and CF$_4^{2+}$ are calculated using the restricted active space self-consistent-field (RASSCF) \cite{Werner1981,Malmqvist1990} and restricted active space perturbation theory (RASPT2) methods \cite{Finley1998,Malmqvist2008}, and are followed by the calculation of transition dipole moments between the ground/valence state and core-excited state by RASSI module in OpenMolcas suite \cite{Aquilante2020}. 
The vibrational broadening of the neutral CF$_4$ molecule (Fig.~\ref{fig:edge}) was taken into account by displacing the equilibrium geometry along the mostly IR active umbrella motion.
For dissociating molecules CF$_4^+$ and CF$_4^{2+}$, induced by photoionization, we optimized 
the planar radicals CF$_3^+$ and CF$_3^{2+}$ respectively, where the dissociated neutral F is displaced at 5 \AA~from the central carbon.
Then, the geometrical difference between the dissociated molecule and the equilibrium geometry at the ground state is linearly discretized by 8 numerical grids. The electronic structure calculations at different charge states (CF$_4$, CF$_4^+$, and CF$_4^{2+}$) were carried out at 8 different geometries, and their absorption spectra were averaged to obtain Fig.~\ref{fig:pre-edge}d and ~\ref{fig:edge}c.
The basis function is ANO-RCC-VTZP\cite{Roos_main_2004}.
The active space is the same for the three molecular species, which is designed to implement the projection technique for the calculation of the highly excited states~\cite{Delcey2019}.  
The four F-1s orbitals are put in the RAS1 space, allowing for one hole at most. The RAS2 space consists of 11 orbitals, i.e., for CF$_4$ and CF$_4^+$, there are 5 highest occupied orbitals and 6 lowest unoccupied orbitals, while for CF$_4^{2+}$, there are 4 highest occupied orbitals and 7 lowest unoccupied orbitals.
Upper states (with a single core-hole) of the transitions are optimized separately with the same active space, but restricted to configurations with one hole in RAS1. 
Following the RASSCF calculation, the multi-state RASPT2 method is employed to include dynamical correlations~\cite{Sauri2011} for both the lower and upper states of the transitions. An imaginary shift of 0.2 a.u. (developer recommended value) is used to the external part of the zeroth-order Hamiltonian to eliminate intruder states~\cite{Forsberg1997}.

\section{Molecular dynamics simulations}\label{app:ras-md}
\FloatBarrier
\begin{figure}
	\includegraphics[width=\figurewidth]{{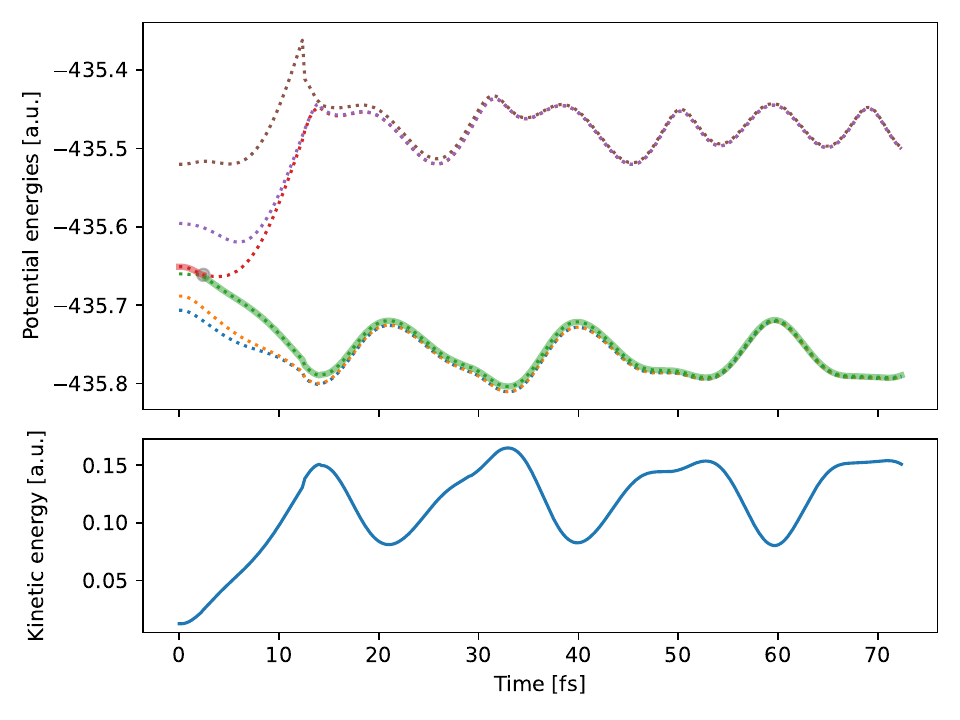}}
	\caption{\label{fig:hop} Molecular dissociation dynamics starting from the fourth valence-excited state of CF$_4^+$. (a) The potential energies of the relevant 6 states obtained from RASSCF during the propagation. Surface hopping, indicated by the gray circle, from state $\left|4\right>$ (red solid curve) to state $\left|3\right>$ (green solid curve) enables the dissociation process. (b) Total kinetic energy of all nuclei. }
\end{figure}
The fragmentation dynamics are modeled using the semiclassical surface hopping molecular dynamics module DYNAMIX in OpenMolcas~\cite{Aquilante2020}, based on the CASSCF energy surfaces. 
The active space for CF$_4^{+}$ is CAS(11;10), consisting of configurations permuting 11 electrons in 6 highest occupied (4t$_2$ and 1t$_1$) and 4 lowest unoccupied orbitals (5t$_2$ and 5a$_1$).
For CF$_4^{+2}$ the selection of active-space is CAS(10,10), consisting of configurations permuting 10 electrons in 5 highest occupied (4t$_2$ and 1t$_1$) and 5 lowest unoccupied orbitals (1t$_1$, 5t$_2$ and 5a$_1$).
To study the dissociation dynamics for different ionization products with various valence-hole distribution possibilities, six energy surfaces are calculated. We carry out six separate sets of MD simulations where molecules are initially put on a different energy surface.
Note that the choice of these states is solely based on the feasibility of calculation; they may not correspond to the most populated final states of photoionization or Auger-Meitner ionization.
Surface hopping is used to account for nonadiabatic transitions at possible conical intersections during the path of dissociation~\cite{Tully1990,Hammes‐Schiffer1994}. Decoherence correction with a typical value of 0.1 Hartree is used in the simulations~\cite{Granucci2007}.

Each set of simulations starting from a given molecular electronic energy state further consists of an ensemble of five parallel calculation trajectories of single molecules.
The initial velocities of each nuclear constituent are set as a random value sampled with a Maxwell-Boltzmann distribution centered at the temperature of 300 K.
During the simulation, the velocities of nuclei are rescaled to guarantee the conservation of total energy. 
The simulations starting with different velocities and initial energy surfaces show similar dissociation dynamics for CF$_4^+$ as highlighted in Fig.~\ref{fig:RASMD}(a) and (d), while two representitive dissociation patterns are found for CF$_4^{2+}$ as highlighted in Fig.~\ref{fig:RASMD}(b), (e) as well as (c) and (f). 
The molecular dissociation dynamics starting from the fourth state is shown in Fig.~\ref{fig:hop} as an example of the effect of surface hopping. 
In this case, the surface hopping occurs at the early stage of the propagation (at 2.5 fs), the switch from the  $\left|4\right>$ (red solid curve) to the state $\left|3\right>$ (green solid curve) enables the dissociation of the molecule.

\normalem
\providecommand{\noopsort}[1]{}\providecommand{\singleletter}[1]{#1}%

\end{document}